\documentclass{article}
\usepackage{amsfonts,amssymb, amsmath}

\def\bq{ \begin{equation}}
\def\eq{ \end{equation}}
\def\ben{ \begin{eqnarray}}
\def\en{ \end{eqnarray}}

\begin{document}


\title{On computation of Darboux polynomials for full Toda lattice}
\author{A.V. Tsiganov\\
\it\small St.Petersburg State University, St.Petersburg, Russia\\
\it\small e--mail: andrey.tsiganov@gmail.com}

\date{}
\maketitle

\begin{abstract}
One of the oldest methods for computing invariants of ordinary differential equations is tested using the full Toda lattice model. We show that the standard method of undetermined coefficients and modern symbolic algebra tools together with sufficient computing power allow to compute Darboux invariants without any additional information.
 \end{abstract}

\section{Introduction}
\setcounter{equation}{0}
A numerical simulation of differential equations is an essential component of a variety of applications.
The fundamental properties of computational methods, such as accuracy, stability, convergence, and computational efficiency, are considered to be of great importance when determining the utility of a numerical algorithm. In recent developments, various aspects of geometric structure preservation have come to the fore as a significant addition to these fundamental properties  \cite{book0,book1,book2}.

Symbolic AI, which translates implicit human knowledge into a more formalised and declarative form based on rules and logic, should automatically develop its own structure-preserving method for each system of differential equations under consideration. For example, for the Kepler problem, the first integrators conserved the energy and angular momentum vector. Modern methods also preserve the Runge-Lenz vector as well as the volume form. The objective at this time is not to select the most suitable existing integrator for a given problem; rather, it is to construct the most efficient integrator for a given problem.

In the context of structure-preserving numerical methods, it is not essential for the first integrals of the Kepler problem to commute with each other. What is crucial is that the discrete trajectories lie on the intersection of the invariant algebraic varieties, and that discretization is a measure preserving map, because the measure-preserving discrete maps are easier to control using ergodic theory methods \cite{erg01}. 

We aim to discuss a method of undetermined coefficients appearing in the Darboux construction of invariant algebraic varieties and invariant volume forms \cite{d78,d78a}. In the Darboux theory we do not use Lagrangian or Hamiltonian form of equations, symplectic or Poisson geometry, qualitative and asymptotic analysis of differential equations, Lax matrices, representation theory, etc.  We start with  a given vector field and  solve algebraic equations which are the backbone of many machine learning algorithms and techniques and, therefore, now we have many modern computer implementations of the desired  computational methods. 

As a benchmark model we chose the full Toda lattice defined by the Lax equations, which encodes all invariant structures and allows us to compare the obtained results with known ones.  The Toda systems are a large class of integrable systems for which the entries are homogeneous and inhomogeneous polynomials of second order. This means that the corresponding algebraic equations are linear.

There are few  methods to compute Darboux polynomials for a polynomial vector field including 
method of undetermined coefficients, method of factorisation of polynomials and the Kowalevski-Painlev\'{e} method.
In this note, we demonstrate that method of undetermined coefficients can be more efficient in practice than other methods.

\section{Darboux polynomials and invariant algebraic varieties}

In 1878, Darboux provided a new theory of the integrability of ordinary differential equations, based on the existence of invariant algebraic varieties and invariant volume forms defined by Darboux polynomials \cite{d78,d78a}. 
In 1891 his work was continued by Poincar\'{e} \cite{poi91}, Painlev\'{e} \cite{pain91} and Autonne \cite{ann91}.
Now there exist a lot of different names in the literature for Darboux polynomials, for example we can find:
special integrals, eigenpolynomials, algebraic invariant curves, invariant algebraic hypersurfaces, particular algebraic
solutions, special polynomials, semi invariants, etc.

Let us consider a system of  differential equations  
\bq\label{m-eq}
\frac{dx_1}{X_1}= \frac{dx_2}{X_2}=\cdots=\frac{dx_n}{X_n}
\eq 
where entries of the vector field  $X_k$ are polynomials of degree at most $d$ on $x$. We say that $P$ is a Darboux polynomial of $X$  if there exists a polynomial $c$ of degree at most $d-1$, called the cofactor of $P$,  such that
\bq\label{d-pol}
\partial P(x)  =c(x)P(x)\,,\qquad \partial=\sum_{i=1}^n X_i\dfrac{\partial  }{\partial x_i}\,.
\eq
Using a set of  Darboux polynomials
\begin{equation*}
\partial  {P}_m(x) =  c_m(x) {P}_m(x), \qquad m=1,\dots,\ell, 
\end{equation*}
we can define solution of (\ref{d-pol}) depending on arbitrary parameters $\alpha_m$
\begin{equation}\label{d-comb}
P(x)=\prod_{m=1}^{\ell}  {P}_m^{\alpha_m }(x)\,,\qquad\mbox{where}\qquad  c(x)=\sum_{m=1}^{\ell} \alpha_m  c_m(x)\,.
\end{equation}
If new cofactor  is equal to  zero 
\[
c(x)=\alpha_1  c_1(x)+\alpha_2  c_2(x)+\cdots \alpha_\ell  c_\ell(x)=0
\]
then $P$ (\ref{d-comb}) is a first integral of $X$.  If cofactor is a non-zero constant $c=C$,  then 
\[\partial \ln P(x)=C\,,\qquad P\left( {x}\left(t\right)\right) = P\left( {x}\left(0\right)\right) e^{C t},\] 
and Darboux polynomial $P$ (\ref{d-comb}) defines an invariant foliation of the state space.

If new cofactor is equal to divergence of $X$
\bq\label{div-c}
c(x)=\alpha_1  c_1(x)+\alpha_2  c_2(x)+\cdots \alpha_\ell  c_\ell(x)=\mbox{div} X\,,
\eq
then $P(x)$ (\ref{d-comb}) is a Jacobi multiplier $M$ \cite{jac-book} which satisfies to equation 
\bq\label{j-eq}
\mbox{div} (MX)=\sum_{i=1}^n \frac{\partial (MX_i)}{\partial x_i}=0\,,\qquad M=P(x)\,.
\eq
It allows us to define a volume form 
\[
\Omega=\mu\, d x_1\wedge \cdots\wedge dx_n\,,\qquad \mu=M^{-1}\,,
\]
which is invariant to flow of vector field $X$.  Recall that for a measure-preserving system, the Poincar\'{e} recurrence theorem states that almost all points of the state space are infinitely recurrent with respect to $X$. We can control the discretization procedure using this and other theorems from the ergodic theory, see \cite{erg01}.

From (\ref{d-pol}) it follows that if a solution curve of system (\ref{m-eq}) has a point on the   algebraic
variety defined by Darboux polynomial $P$
\[P(x_1,\ldots,x_n)=0 \,,\] 
then the whole solution curve is also contained in  this algebraic variety which is invariant by the flow of the system. The invariant algebraic varieties  $P_1(x)=0,\ldots,P_\ell(x)=0$ divide the state space into invariant parts, which makes it easier to study the dynamics of the vector field $X$.

\subsection{Computation of the Darboux polynomials}
The most direct way to compute Darboux polynomials is the method of undetermined coefficients. In other words, we substitute into (\ref{d-pol}) the polynomials $P$ and $c$ with the unknown coefficients and solve the corresponding system of polynomial equations for these coefficients.

For the full Toda lattice $X_i$ are polynomials of second order, 
so that cofactors $c_1(x),\ldots,c_\ell(x)$ are linear  polynomials so that 
\[
\mbox{div}X=\alpha_1 c_1(x)+\ldots+\alpha_\ell c_\ell(x)\,,\qquad \alpha_m\in\mathbb N_+\,.
\]
After a convenient linear change of coordinates $c_1,\ldots,c_\ell $ becomes $x_1,\ldots, x_\ell$. Substituting known cofactor $c_m$  into  (\ref{d-pol}) we can compute Darboux polynomials $P_m^{(j)}$ of degree $j$ solving a linear system of algebraic equations on the undetermined coefficients. 

In 1911  Lagutinskii proposed to compute cofactors $c(x)$ using critical points of $X$ associated with singular solutions obtained in framework of the Kowalevski-Painlev\'{e} analysis, see section 4 in  \cite{lag1} and discussion in \cite{gor01,ll17}. Substituting
formal Laurent series
\[
x_i=\sum_{k=0}^\infty a_{ik}t^{k+\beta_i}
\]
into (\ref{m-eq})  we can compute a dominant balance $\beta_i$ and $a_{i0}$
and recursively find other coefficients $a_{ik}$. Then, substituting these data into $\partial P(x)=c(x)P(x)$ (\ref{d-pol})   we obtain a weight-homogeneous component of the cofactor polynomial $c(x)$ of highest weighted degree, see Theorem 5.7 in \cite{gor01} and details in \cite{ll17}.

One computational problem  is that we use the field of complex numbers within the Kowalevski-Painlev\'{e} analysis. The second problem is that for full  Toda lattices the number of solutions of equations 
\[x_iX_j-x_iX_j=0\,,\qquad i,j=1,\ldots,n\,,\]
is too large and, according to \cite{lag1}, we need only in a subset of  these solutions. Devoting resources to finding all the solutions is therefore not rational.

In \cite{lag1,lag2} Lagutinskii proposed to compute Darboux polynomials $P(x)$ of degree $N$ by using factorization of polynomial
\begin{equation}\label{m-lag}
W=\mbox{det}\,\left(
\begin{array}{cccc}
B_{1} & B_{2} & \ldots & B_{s} \\
\partial B_{1} & \partial B_{2} & \ldots & \partial B_{s} \\
\vdots & \vdots & \ldots & \vdots \\
\partial ^{s-1}B_{1} & \partial ^{s-1}B_{2} & \ldots & \partial ^{s-1}B_{s}
\end{array}
\right)\,.
\end{equation}
Here polynomials $B_j$ are all the monomials on $x_1,\ldots,x_n$ of fixed degree $N$. In modern literatures $W$ is so-called  extactic polynomials of $X$, see \cite{ll07,ll09,ll12,ch11}. 

Of course, we know efficient modern methods for factorization, but the main problem for the Toda lattice is computing derivatives of monomials and the corresponding determinants. As an example, let us take the full symmetric $sl(4)$ Toda lattice and try to compute Darboux polynomials of the second order in this way. In this case $n=10$ and $s=55$, so we have to compute derivatives of order $54$ on $55$ monomials $B_{ij}=x_ix_j$, then compute $55\times 55$ determinant and its factorization. In practice, we can do this by spending considerably more time than with the method of undetermined coefficients.

 \section{Full symmetric Toda system}
 
According to \cite{k79,s80,dei86} full Toda lattice is defined as the coadjoint action of a Borel
subgroup of a simple Lie group  on the dual Lie algebra of this Borel subgroup. Different Cartan decompositions 
correspond to several different versions of Toda lattice.

In this section a more elementary approach is employed, whereby a full symmetric Toda system is defined by means of an $N\times N$ symmetric matrix $L$
\[
L=\begin{pmatrix}
       x_{1,1} & x_{1,2} & x_{1,3} &  \dots & x_{1,N}\\
       x_{1,2} & x_{2,2} & x_{2,3} &  \dots & x_{2,N}\\
       x_{1,3}  & x_{2,3} & x_{3,3} & \dots & x_{3,N}\\
       \vdots &\vdots &  & \ddots & \vdots\\
       x_{1,N} & x_{2,N} & \dots      & x_{N-1,N} & x_{N,N}
    \end{pmatrix},
\]
and antisymmetric matrix
\[    
    B=\begin{pmatrix}
       0 & x_{1,2} & x_{1,3} &  \dots & x_{1,N}\\
       -x_{1,2} & 0 & x_{2,3} &  \dots & x_{2,N}\\
       -x_{1,3}  & -x_{2,3} & 0 & \dots & x_{3,N}\\
       \vdots &\vdots &  & \ddots & \vdots\\
       -x_{1,N} & -x_{2,N} & \dots & -x_{N-1,N} & 0
    \end{pmatrix}.
\]
The $n=N(N+1)/2$ coordinates  
\[
x=\{ x_{1,1},x_{1,2},\ldots,x_{N-1,N},x_{N,N} \}
\]
satisfy to the system of differential equations (\ref{m-eq}) with the vector field $X$
defined by the Lax equation
\bq\label{lax-eq}
\frac{d}{dt}L=BL-LB\,.
\eq
Entries of the vector field $X$ in (\ref{m-eq}) 
\[
X_{i}=\frac{d}{dt}\, x_i\,,\qquad i=1,\ldots,n\,,
\] 
are homogeneous polynomials of second order in coordinates $x$. 

\subsection{The method of undetermined coefficients}
A set  of $n$ pairs $(x_i,X_i)$ is a starting point for calculation of Darboux polynomials.
 In the first step of algorithm we compute divergency of $X$ 
\[
c(x)=\mbox{div} X=\sum_{k=1}^N \alpha_kx_{k,k}\,,
\]
where $\alpha_k$ are some integer numbers and $x_{k,k}$ are diagonal elements of the Lax matrix. 
We suppose that it is a sum of cofactors which are linear polynomials $c_m(x)$ in $x_{k,k}$ and cofactor $c_0=0$.

In the second step we take Darboux cofactor $c_0=0$ and consider equations
\bq\label{0-cofactor}
\partial P_0^{(j)}(x)=0\,,
\eq
on a ring of formal power series $\mathbb C[[x_1,\ldots, x_n]]$ endowed with a Lie  derivative along vector field X
\[
\partial=\sum_{i=1}^n X_i\dfrac{\partial  }{\partial x_i}.
\] 
Here $P_0^{(j)}(x)$ is a polynomial of degree $j=1,2,\ldots$ with undetermined coefficients. We solve these equations (\ref{0-cofactor}) until we get a solution $P_0^{(j+1)}(x)$ that is linearly dependent on the previous ones $P_0^{(1)}(x),\ldots, P_0^{(j)}(x)$. 

In the third step, we take two linear polynomials with undetermined coefficients  
\[
c_1(x)=\sum_{k=1}^N v_k x_{k,k}\qquad\mbox{and}\qquad P_1^{(1)}(x)=\sum_{i=1}^n u_i x_{i}\,.
\]
First polynomial $c_1(x)$ depends only on diagonal elements of the Lax matrix, whereas second polynomial $P_1^{(1)}(x)$ depends on all the coordinates. Then we compute $c_1(x)$ and $P_1^{(1)}(x)$ solving system of bi-linear equations that is derived from (\ref{d-pol}) which in this case reads as
\bq\label{lin-eq}
\sum_{i=1}^n u_iX_i=\left(\sum_{k=1}^Nv_kx_{k,k}\right)\left(\sum_{i=1}^n u_i x_{i}\right)\,,
\qquad u_{i},v_k\in\mathbb R\,.
\eq
Substitute the cofactor $c_1(x)$ thus found into the equations
\[ 
\partial P_1^{(j)}(x)=c_1(x)P_1^{(j)}(x)\,,\qquad \mbox{deg} P_1^{(j)}(x)>1\,,
\]
where $P_1^{(j)}(x)$ are polynomials of degree $j$ with undetermined coefficients, and obtain systems of linear equations which are directly solved. We calculate $P_1^{(j)}(x)$ until we get a solution that is linearly dependent on the previous ones.

In the fourth step, we take linear polynomial
\[
c_2(x)=\sum_{k=1}^N v_k x_{k,k}\,,\qquad c_2(x)\neq 2c_1(x)\,,
\]
and polynomial of second order 
\[
P_2^{(2)}(x)=\sum_{i,j=1}^n u_{ij} x_{i}x_{j}\,.
\]
We compute $c_2(x)$ and $P_2^{(2)}$ solving system of bi-linear equations  derived from definition (\ref{d-pol}) which now looks like 
\bq\label{sec-eq}
\sum_{i,j=1}^{n} u_{i,j}\left(x_iX_j+x_jX_i\right)= \left(\sum_{k=1}^N v_kx_{k,k}\right)\left(\sum_{i,j=1}^{n} u_{i,j} x_{i}x_{j}\right)\,,\qquad u_{i,j},v_k\in\mathbb R\,.
\eq
Then we substitute obtained cofactor $c_2(x)$ into equations
\[ 
\partial P_2^{(j)}(x)=c_2(x)P_2^{(j)}(x)\,,
\]
where $P_2^{(j)}(x)$ are polynomial of degree $j>2$ with undetermined coefficients. We stop  calculations of polynomials $P_2^{(j)}(x)$ when obtain linear dependent solutions of these linear equations.

In a similar manner, the calculation of cofactors $c_3(x),\ldots,c_\ell(x)$ is performed until the desired result is obtained
\[
\mbox{div}X=c_1(x)+c_2(x)+\ldots+c_\ell(x)\,.
\]
Solutions of arising bi-linear and linear algebraic  equations are directly computed using modern AI software for a  polynomial time. For instance, for $N=10$, all of these calculations on a standard laptop take less than 10 minutes. The majority of the time is spent finding the irreducible Darboux polynomials among the set of obtained solutions.

In the proposed strategy of computation, we have hidden some implicit human knowledge. First, we  used homogeneity of $X$ and definition of the Darboux polynomials to prove that $P_m^{(j)}(x)$ are homogeneous polynomials of order $j$ in $x$. Second, we computed the divergence div$X$ and then assumed that all cofactors $c_m(x)$ are linear polynomials of $N$ diagonal elements 
\[
c_m(x)=v_1x_{1,1}+v_2x_{2,2}+\cdots+v_Nx_{N,N}\,.
\]   
This has simplified our calculations, since in generic case we have to use linear polynomials of all variables 
\[
c_m(x)=v_1x_{1}+v_2x_{2}+\cdots+v_nx_{n}\,,\qquad n=N(N+1)/2\,,
\]  
and inhomogeneous Darboux polynomials with much more undetermined coefficients. 

So, in the framework of symbolic AI, we need to extract this implicit information directly from the input data. As an example, we can substitute generic linear polynomials and generic inhomogeneous Darboux polynomials into the equation $\partial P(x)=c(x)P(x)$ (\ref{d-pol}) and solve the obtained equations at $N=3,4,5,6$, which is quite possible today. We can then try to train our algorithm on this data. It allows us to extract and formalize the information we need for calculations at $N>6$.

\subsection{Comparison with known results}
Let us express obtained results in terms of the Lax matrix in order to make a comparison with the results that are already known. For odd $N$ divergency of $X$ reads as
\[\mbox{div}\,X=2\sum_{m=1}^\ell c_m(x)\,,\qquad\ell=\frac{N-1}{2}\,,
\]
for even $N$ it looks like
\[\mbox{div}\,X=2\sum_{m=1}^{\ell-1} c_m(x)+c_\ell(x)\,,\qquad \ell=\frac{N}{2}\,,
\]
where cofactors $c_1(x),\ldots,c_\ell(x)$ are equal to 
\[
c_m(x)=\sum_{k=1}^m x_{k',k'}-\sum_{k=1}^m x_{k,k}\,,\qquad k'=N+1-k\,, \qquad m=1,\cdots,\ell\,.
\]
Using $c_0=0$ we obtain $n$ solutions of the equation (\ref{0-cofactor})
\bq\label{p0-symm}
P_0^{(j)}(x)=\frac{1}{2j}\,\mbox{trace} L^j\,,\qquad j=1,\ldots,n\,.
\eq
It is well known fact directly follows from the Lax equation (\ref{lax-eq}). We reproduced it without using the Lax matrices which are only used to formalize the obtained result.  

Substituting non-trivial cofactors $c_m(x)\neq 0$ into
\bq\label{k-cofactor}
\partial P^{(j)}_m(x)=c_m(x)P_m^{(j)}(x)\,,\qquad m=1,\ldots,\ell
\eq
we again obtain a system of linear algebraic equations on  undetermined coefficients.  As above we formalize the obtained result using Lax matrices and present  only irreducible Darboux polynomials 
\[
P_m^{(j)}(x)=\mbox{det}\, A_m\,,\qquad A=L^j\,,\qquad j=1,\ldots,n\,.
\]  
Here $A_m$ is defined as an $(n-m)\times (n-m)$ matrix derived from $n\times n$ matrix $A$ through the elimination of columns $1$ through $m$-th and rows $m$-th through $n$-th. These polynomials are minors of $L^j$, lying on the intersection of columns and rows associated with the cofactor $c_m(x)$  involving diagonal entries only.

So, we  reproduced known result from \cite{ch15}  without using the Lax matrices.
Using non-zero Darboux polynomials  from this set we can define rational first integrals
\[
J_m^{(k_1,k_2)}=\frac{ P_m^{(k_1)}(x) }{P_m^{(k_2)}(x)}
\]
 and Jacobi multipliers for even and for odd $N$
\[
M_{odd}^{(j)}=\left(\prod_{m=1}^\ell P_m^{(j)}\right)^2\,,\qquad
M_{even}^{(j)}=\left(\prod_{m=1}^{\ell-1} P_m^{(j)}\right)^2 \cdot P^{(j)}_\ell\,.
\]
Other solutions of Jacobi's equation (\ref{j-eq}) are obtained by multiplying these polynomials by first integrals of $X$.
The corresponding invariant measure $\mu=M^{-1}$ defines invariant volume form $\Omega$ of  the given vector field $X$.

So, for full symmetric Toda lattice we have $n-2$ independent first integrals and an invariant volume form and, therefore, this system is integrable in quadratures according to the Euler-Jacobi last multiplier theorem \cite{jac-book}.  

The structure of such dynamical systems in a neighborhood of compact integral manifolds without singular points is described, for example, \cite{koz93}. It is based on the classical results due to Poincar\'{e}, Siegel and Kholmogorov concerning the dynamical systems on a two dimensional torus. This information must be taken into account in the construction of numerical schemes preserving volume form $\Omega$ and invariant algebraic varieties defined by 
\[P_m^{(j)}(x)=0\,.\] 

\subsection{Examples}
At $N=4$, input consists of  $n=N(N+1)/2=10$  pairs $(x_i,X_i)$.
Divergency of the vector field $X$ is equal to
 \[
\mbox{div}X=-3 x_{1,1}-x_{2,2}+x_{3,3}+3 x_{4,4}\,.
\]
Substituting first cofactor $c_0=0$  from this sum into (\ref{0-cofactor}) we obtain the system of  linear  equations
\[
\partial \sum u_{i_1\ldots i_j}x_{i_1}\cdots x_{i_j}=0\,,
\]
for undetermined coefficients $u_{i_1\ldots i_j}$. Solving these equations we obtain four independent irreducible Darboux polynomials $P_0^{(1)}(x),\ldots, P_0^{(4)}(x)$ which we will not present explicitly. Of course they are equal to polynomials (\ref{p0-symm}) up to a scaling factor.

Solving bi-linear equations (\ref{lin-eq}) we obtain
\[c_1(x)=x_{4,4}-x_{1,1}\,.\]
Substituting this cofactor into  (\ref{k-cofactor}) we get  systems of linear equations on undetermined coefficients. Irreducible solutions of these equations are
\begin{align*}
P_1^{(1)}(x)=&x_{1,4}\,,\qquad 
P_1^{(2)}(x)=x_{1,2} x_{2,4}+x_{3,4} x_{1,3}-x_{1,4} x_{2,2}-x_{1,4} x_{3,3}\,,\\
P_1^{(3)}(x)=&
\left(x_{1,1} x_{1,2}+x_{1,2} x_{2,2}+x_{4,4} x_{1,2}+x_{1,3} x_{2,3}\right) x_{2,4}+
\left(x_{1,1} x_{1,3}+x_{1,2} x_{2,3}+x_{1,3} x_{3,3}\right)x_{3,4} \\
+&\left(x_{4,4} x_{1,1}-x_{2,3}^{2}\right) x_{1,4}+x_{3,4} x_{1,3}x_{4,4}+\frac{\left(x_{1,1}^{2}-x_{2,2}^{2}-x_{3,3}^{2}+x_{4,4}^{2}\right) x_{1,4}}{2}\,.
\end{align*}
Solution  $P_1^{(4)}$ is the polynomial of fourth order which dependents on the previous Darboux polynomials.

So, we have to take the next step and solve a system of bi-linear equations involving next cofactor $c_2\neq 2c_1$ (\ref{sec-eq}). In this case solution is
and obtain 
\[
 c_2(x)=\sum_{k=1}^4 v_kx_{k,k}=x_{3,3}+x_{4,4}-x_{1,1}-x_{2,2}\,,\qquad
P_2^{(2)}(x)=x_{13}x_{24} - x_{1 4}x_{2 3}\,.
\]
Substituting $c_2(x)$ into (\ref{k-cofactor}) and solving the resulting linear systems of equations
we obtain the Darboux polynomials of the third and fourth order $P_2^{(3)}(x)$ and $P_2^{(4)}(x)$, which we will also omit for the sake of brevity. 

Since 
\[
\mbox{div}X=-3 x_{1,1}-x_{2,2}+x_{3,3}+3 x_{4,4}=2c_1(x)+c_2(x)\,,
\]
we can stop the calculations.

Output consists of three cofactors $c_0(x),c_1(x),c_2(x)$ and 10 linearly independent Darboux polynomials $P_m^{(j)}(x)$, which were obtained directly from differential equations (\ref{m-eq}). All these calculations take seconds on a standard laptop using standard  \texttt{Maple} solvers. Using specialized software, we obtained analogous results for the case $N=20$ in a few minutes.

Let us briefly discuss also case of the odd $N=7$ when input consists of $n=28$ pairs $(x_i,X_i)$.  Equation (\ref{lin-eq}) gives rise to 226 bi-linear equations with the following solution
\[c_1(x)=c_0(x)+x_{7,7}-x_{1,1}\,,\qquad P_1^{(1)}(x)= x_{1,7}\,,\qquad c_0(x)=0\,.\]
Second system of bi-linear equations  (\ref{sec-eq}) consists of 2128 equations which have solution
\[
c_2(x)=c_1(x)+x_{6,6}-x_{2,2}\,,\qquad P_2^{(2)}(x)=x_{1 ,6}x_{2,7} - x_{1,7}x_{2,6}\,.
\]  
Similar system for third cofactor consists of 24636 equations  that take about a minute to solve using standard solver in \texttt{Maple}. Solution reads as  
\[
c_3(x)=c_2(x)+x_{,55}-x_{3,3}\,,\]
and
\[P_3^{(3)}(x)=x_{1 ,5} x_{2, 6} x_{3, 7}-x_{1, 5} x_{2, 7} x_{3, 6}-x_{1, 6} x_{2, 5} x_{3, 7}+x_{1, 6} x_{2, 7} x_{3, 5}+x_{1, 7} x_{2 ,5} x_{3, 6}-x_{1, 7} x_{2, 6} x_{3, 5}\,.
\]
Linear systems of equations for other polynomials $P_m^{(j)}(x)$ of $j>m$ degree are more effectively solved in  \texttt{Maple} using software developed specially for solution of linear equations.

\section{Full Kostant-Toda lattice}
Following \cite{sin91} now we consider the Lax equation (\ref{lax-eq}) not for symmetric matrices, but
for Hessenberg matrices
\[
L=\begin{pmatrix}
       * & 1 & 0 &  \cdots & 0\\
       * & * & 1 &  \ddots & 0\\
       * &\ddots & \ddots & \ddots&0\\
       \vdots&\ddots & *&* & 1\\
       * & \cdots &  * & * & *
    \end{pmatrix}.
\]
In this case equations of motion have the form 
\bq\label{kt-eq}
\frac{d a_i}{d t} = b_i \qquad\mbox{and}\qquad \frac{d b_i}{d t} =  \sum A_{ij}\,a_ib_j + \sum B_jb_j \,.
\eq
Here $a_i$ are diagonal elements of the Lax matrix,  $b_i$ are other non-constant elements of the Lax matrix and numerical parameters $A_{ij}$ and $B_j$ are defined by a complex simple Lie algebra $\mathfrak g$, its Cartan subalgebra $\mathfrak h$ and a chosen set of positive roots of $\mathfrak g$. 

In this case components of the corresponding vector field $X_i=\dot{x}_i$ are inhomogeneous polynomials of second order in $x=(a,b)$ and divergency of $X$ is a linear polynomial on diagonal elements $a_i$
\[
\mbox{div} X= \alpha_1 a_{1}+\cdots+\alpha_\ell a_{\ell}\,.
\]
Substituting  formal Laurent series into equations of motion we obtain 
\[
a_{i}(t) = \frac{1}{t}\sum \limits_{k = 0}^{\infty} a_{ik}t^k, \qquad b_{j}(t) = \frac{1}{t^{\beta_j}} \sum \limits_{k = 0}^{\infty}b_{j k}t^k\,,\qquad \beta_j>1\,,
\]
where $\beta_j$ and coefficients $a_{ik}$ and $b_{jk}$ can be directly computed, see \cite{sin91} and \cite{xie22}. So, using Theorem 5.7 in \cite{gor01} and computations in \cite{ll17} we can prove that all the cofactors are linear polynomials on diagonal elements only
\[
c_m(x)= \alpha_{1,m} a_{1}+\cdots+\alpha_{\ell,m} a_{\ell}\,.
\]
For the full symmetric Toda lattice, all components of the vector field $X$ are homogeneous polynomials of second order, and thus the cofactors could be linear polynomials on all variables $x_i$, at least until we can prove otherwise.

The homogeneity of the vector field for full symmetric Toda lattice implies that the Darboux polynomials are also homogeneous according to (\ref{d-pol}). For the full Kostant-Toda lattice, the vector field is inhomogeneous and the desired Darboux polynomials have the form
\[
\partial P_m^{(j)}(x)=c_m(x) P_m^{(j)}(x)\,,\qquad P_m^{(j)}(x)=Q_m^{(j)}(x)+R_m^{(j-1)}(x)\,,\qquad 1\leq m \leq \ell\,,
\]
where $Q_m^{(j)}(x)$ and $R_m^{(j-1)}(x)$ are homogeneous polynomials of degree $j$ and $j-1$, respectively.

As a benchmark model we chose the full Kostant-Toda lattice on the Lie algebra $so(5)$ since the Laurent series solutions for this model readers can found in \cite{xie22}. It allows us to discuss some ideas from the Lagutinskii papers \cite{lag1,lag2} and compare the corresponding computations with the method of undetermined coefficients.   

\subsection{Example}
In this section we follow to \cite{xie22} where all the definitions and details can be found. The Lax equation for full Kostant-Toda lattice is 
\[
\frac{d}{dt} L=[B,L]\,,
\]
where
\[
L=\left(\begin{smallmatrix}
a_2& 1& 0& 0& 0\\ b_2& 2 a_1 - a_2& 1& 0& 0\\ 2 c_1& 2 b_1& 0& 1& 0\\ 4 d_1& 0& 2 b_1& a_2 - 2 a_1& 1\\ 0& 4 d_1& -2 c_1& b_2& -a_2
\end{smallmatrix}\right)\,,\qquad
B=\left(\begin{smallmatrix}
a_2& 1& 0& 0& 0\\ 0& 2 a_1 - a_2& 1& 0& 0\\ 0& 0& 0& 1& 0\\ 0& 0& 0& a_2 - 2 a_1& 1\\ 0& 0& 0&0& -a_2
\end{smallmatrix}\right)
\]
Here $x=(a_1,a_2,b_1,b_2,c_1,d_1)$ are coordinates on a six-dimensional state space for which we are using the notation $a_{1,2},b_{1,2},c_1$ and $d_1$ from the paper \cite{xie22}. To avoid confusion, the cofactor polynomials are denoted $c_m(x)$.

The entries of vector field $X$ in (\ref{m-eq}) are inhomogeneous polynomials of second order
\bq\label{x-toda}
X=\left(\begin{array}{c}
b_1\\ b_2\\(a_2 - 2a_1)b_1 - c_1 \\(2a_1 - 2a_2)b_2 + 2c_1\\-a_2c_1 + 2d_1\\-2a_1d_1
\end{array}
\right)\,,
\eq
so cofactor polynomials $c_m(x)$ are linear polynomials.

Following to Lagutiskii construction \cite{lag1} we can solve 15 equations 
\[
x_iX_j-x_jX_i=0\,,\qquad i,j=1,\ldots,n=6\,,
\] 
for $x$ and obtain six critical points of the  given vector field $X$
\[\begin{array}{llllll}
\{a_1 = 0,& a_2 = a_2,& b_1 = 0,& b_2 = -2a_2^2,& c_1 = 0,& d_1 = 0\}\\
\{a_1 = a_1, &a_2 = 0, &b_1 = -2a_1^2, &b_2 = 0, &c_1 = 0, &d_1 = 0\}\\
\{a_1 = a_1, &a_2 = a_2, &b_1 = 0, &b_2 = 0, &c_1 = 0, &d_1 = 0\}\\
\{
a_1 = \frac{a_2}3,& a_2 = a_2, &b_1 = -\frac{2a_2^2}{9}, &b_2 = -\frac{2a_2^2}{3}, &c_1 = -\frac{2a_2^3}{9},& d_1 = \frac{-a_2^4}{27}\}\\
\{
a_1 =\frac{3a_2}{4},& a_2 = a_2, &b_1 = -\frac{3a_2^2}{8},& b_2 = -\frac{a_2^2}{2}, &c_1 = 0, &d_1 = 0
\}\\
\{
a_1 = \frac{3 - \sqrt{5}a_2}{4},& a_2 = a_2, &b_1 = -\frac{3 - \sqrt{5}a_2^2}{4},& b_2 = -a_2^2,& c_1 = \frac{1 - \sqrt{5}a_2^3}{4},& d_1 = 0
\}
\end{array}
\]
Substituting these solutions into the linear polynomial with undetermined coefficients
\bq\label{c-fkt}
c(x)=v_1a_1+v_2a_2+v_3b_1+v_4b_2+v_5c_1+v_6d_1
\eq
and consider the linear part of the resulting expression  we obtain desired cofactors. In this case we have
\[
c(x)=v_1 a_1 \qquad\mbox{or}\qquad c(x)=v_2a_2\,.
\]
Recall that if $c(x)$ is the cofactor of the polynomial $P(x)$, then $\alpha c(x)$ is the cofactor of the polynomial $P^\alpha(x)$. Note also that we do not have to calculate all of the six critical points in order to get this result. 

Using eight formal Laurent series solutions obtained in \cite{xie22}, see Table 2, and Theorem 5.7 from \cite{gor01} with a modern complement from \cite{ll17}, we can reproduce the same result for cofactor polynomials in the framework of Kowalevski-Painlev\'{e} analysis.   

In the method of undetermined coefficients we reproduce this result by avoiding some redundant calculations.
Our starting point is the divergence calculation
\[
\mbox{div} X=-2a_1 - 2a_2\,.
\]
Then we take cofactor $c_0(x)=0$ and solve linear equations on undetermined coefficients deriving from  
\[\partial P_0^{(j)}(x)=0\,,\qquad\mbox{where}\qquad j=\mbox{deg}P_0^{(j)}(x)\,. \]
As a result we obtain two independent irreducible Darboux polynomials 
\begin{align*}
P_0^{(2)}(x)=&2a_1^2 - 2a_1a_2 + a_2^2 + 2b_1 + b_2\,,\\ \\
P_0^{(4)}(x)=&a_1^4 - 2a_1^3a_2 + (a_2^2 + 2b_1 + b_2)a_1^2 - 2(a_2b_1 - c_1)a_1 + b_1^2 + 2d_1\,,
\end{align*}
associated  with two values of the Kowalevski exponents $\rho_1=2$ and $\rho_2=4$ (degrees of the Chevalley invariants, see \cite{xie22}). As above we rewrite these Darboux polynomials using spectral invariants of the Lax matrix
\[
P_0^{(2)}(x)=\frac{1}{4}\mbox{trace}\left(L^2\right)\,,\qquad
 P_0^{(4)}(x)=
\frac{1}{16}\mbox{trace}\left(L^4\right)-\frac{1}{64} \mbox{trace}\left(L^2\right)\,.
\]
Then we take generic linear polynomial $c(x)$ (\ref{c-fkt}) and solve bi-linear equations (\ref{lin-eq}). The result is
\[c_1(x)=-2a_1\,,\qquad P_1^{(1)}(x)=d_1\,,\]
Using obtained cofactor $c_1(x)$ we derive linear equations from (\ref{k-cofactor}) and find independent irreducible solution
\[
P_1^{(3)}(x)=2a_1^2d_1 + 2a_1b_1c_1 - 2a_2b_1c_1 - b_1^2b_2 + 4b_1d_1 + c_1^2\,.
\] 
Then we again take the generic linear polynomial $c(x)$ (\ref{c-fkt}) and solve the bi-linear equations (\ref{sec-eq}) to obtain the following two independent irreducible solutions
\[
c_2(x)=-a_2\,,\qquad  P_2^{(2)}(x)=\frac{(2a_1 - a_2)c_1}{2} - \frac{b_1b_2}{2} + d_1
\]
and
\[c_3(x)=-2a_2\,,\qquad P_3^{(2)}(x)=b_2d_1 - \frac{c_1^2}{2}\,.
\]
Since $\mbox{div}X=c_1(x)+c_3(x)$ we stop the calculations. It is easy to see that we have reproduced the information previously obtained from the Lax equation, see  \cite{xie22} and references within.

So, the input consists of six ordinary differential equations  (\ref{m-eq})
\[
\frac{x_1}{X_1}=\cdots=\frac{x_6}{X_6}
\]
whereas  output consists of cofactors $c_0(x),c_1(x),c_2(x),c_3(x)$ and six independent Darboux polynomials 
\[
\{P_0^{(2)}(x),\quad P_0^{(2)}(x),\quad P_1^{(1)}(x),\quad P_1^{(3)}(x),\quad P_2^{(2)}(x),\quad P_3^{(2)}(x)\}
\]
which were obtained directly from the given differential equations. All the calculations take seconds on a standard laptop using standard \texttt{Maple} solvers.

Substituting these irreducible Darboux  polynomials  into (\ref{d-comb}) we compute Jacobi multiplier
\[M=P_1^{(1)}(x)P_3^{(2)}(x)=d_1\left(b_2d_1-\frac{c_1^2}{2}\right)\]
invariant volume form
\[
\Omega=\mu(x)\, dx^1\wedge\cdots\wedge dx^6\,,\qquad \mu(x)=M^{-1}\,,
\]
and two independent rational first integrals 
\[
J_1=\frac{P_1^{(3)}(x)}{P_1^{(1)}(x)}\qquad \mbox{and}\qquad J_2=\frac{\left(P_2^{(2)(x)}\right)^2}{P_3^{(2)}(x)}\,,
\]
which are independent on the polynomial first integrals $P_0^{(2)}$ and $P_0^{(4)}$.

As a result, the vector field $X$ (\ref{x-toda}) on a six-dimensional state space is integrable by quadratures according to the Euler-Jacobi theorem on the last multiplier \cite{jac-book}, since it preserves the volume form $\Omega$ and four independent first integrals. In particular, it means that this full Kostant-Toda system in a neighbourhood of the compact integral manifold is reduced to dynamical system on the two-dimensional torus \cite{koz93}.

The similar result can be obtained for full Kostant-Toda systems associated to Lie algebras of other types with at most  $10\times 10$ Lax matrices, using the {\texttt{Maple}} solver developed for solving linear algebraic equations. 

\section{Conclusion}
The Toda lattice and its generalizations have become the test models for various techniques and philosophies in the theory of integrable systems. This toy model also has a significant impact on many other branches of mathematics.

In this note we use the full Toda lattice model for testing the method of undetermined coefficients in the framework of Darboux integrability theory. We demonstrate the capacity to compute all Darboux invariants without requiring an additional information that would enable the construction of numerical schemes preserving these invariants.

The Kahan-Hirota-Kimura discretization method is a numerical method that was developed for the purpose of integrating quadratic vector fields in any finite dimension. This method has proven to be highly effective due to its ability to generate discrete systems that often possess discrete Darboux invariants \cite{kah19,kah22}. The computation of discrete Darboux invariants arising from the Kahan-Hirota-Kimura discretization of the full Toda lattice promises to be an interesting subject  for research.
 
Systems satisfying the Euler-Jacobi theorem are reduced to dynamical systems on a two\--di\-men\-si\-onal torus. The discretization of the equations of motion on the torus explicitly preserves all the Darboux invariants by definition. It would be interesting to rewrite this discretization in terms of the initial variables that appear in the Lax matrices of the full Toda lattice.

 \subsection{CRediT authorship contribution statement}
Andrey Tsiganov: Writing,  Validation, Methodology, Investigation, Formal analysis, Conceptualization.

\subsection{Funding}
The study was carried out with the financial support of the Ministry of Science and Higher Education of the Russian Federation in the framework of a scientific project under agreement No. 075-15-2024-631.

\subsection{Declaration of competing interest}

The author declares that he have no known competing financial interests or personal relationships that could have appeared to influence the work reported in this paper.

\subsection{Data availability}
No data was used for the research described in the article.

\end{document}